\documentclass[aps,prl,showpacs,twocolumn]{revtex4}
\usepackage{graphicx}
\begin{document}
\newcommand {\be}{\begin{equation}}
\newcommand {\ee}{\end{equation}}
\newcommand {\bea}{\begin{eqnarray}}
\newcommand {\eea}{\end{eqnarray}}
\newcommand {\nn}{\nonumber}


\title{  Penrose Quantum Antiferromagnet }

\author{ A. Jagannathan and A. Szallas}
\affiliation{Laboratoire de Physique des Solides, CNRS-UMR 8502, Universit\'e
Paris-Sud, 91405 Orsay, France }
\author{ Stefan Wessel}
\affiliation{Institut f\"ur Theoretische Physik III, Universit\"at Stuttgart, 70550 Stuttgart, Germany}
\author{ Michel Duneau}
\affiliation{Centre de Physique Th\'eorique, CNRS-UMR 7644, Ecole Polytechnique, 91128 Palaiseau, France}

\date{\today}

\begin{abstract}
The Penrose tiling is a perfectly ordered two dimensional structure with fivefold symmetry and scale invariance under site decimation. Quantum spin models on such a system can be expected to differ significantly from more conventional structures as a result of its special symmetries. In one dimension, for example, aperiodicity can result in distinctive quantum entanglement properties. In this work, we study ground state properties of the spin-1/2 Heisenberg antiferromagnet on the 
Penrose tiling, a model that could also be pertinent for certain three dimensional antiferromagnetic quasicrystals.
We show, using spin wave theory and quantum Monte Carlo simulation, that
the local staggered magnetizations strongly
depend on the local coordination number $z$ and are minimized 
on some sites of five-fold
symmetry. We present a simple explanation for this behavior in terms of Heisenberg stars. Finally we show how best to represent this complex inhomogeneous ground state, using the ``perpendicular space" representation of the tiling.

\end{abstract}
\maketitle

\begin{figure}[t]
\begin{center}
\includegraphics[scale=0.5]{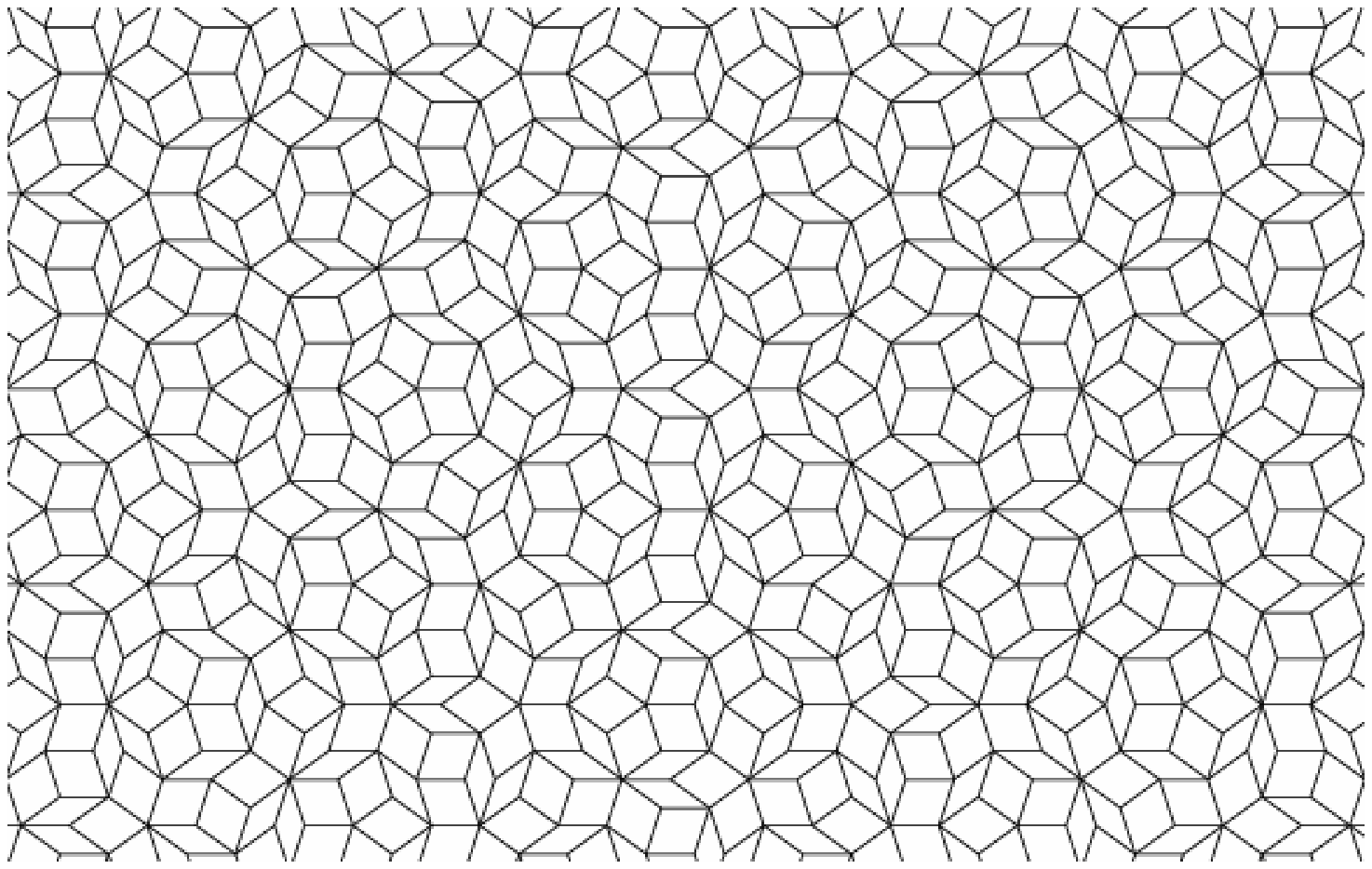}
\vspace{.2cm} \caption {Portion of the Penrose tiling }
\label{fig1.fig}
\end{center}
\end{figure}

The Penrose tiling~\cite{penref}, illustrated in Fig.~\ref{fig1.fig},
is one of the best known of many quasiperiodic tilings.
Its counterpart in one dimension is the Fibonacci chain, while its three dimensional counterpart is the
3D Penrose or icosahedral tiling, the basic template for many quasicrystalline alloys.
One of the most striking and experimentally observable features
of the Penrose tiling is its
five-fold symmetric structure factor with sharp peaks in
reciprocal space. In real space the tiling, built from two types of rhombuses, has a set of vertices of 
coordination number $z$ ranging from 3 to 7, with an overall coordination number of exactly 4. The characteristics of the Penrose tiling such as the tile shapes, or the relative frequencies of vertices can be expressed in terms of the golden mean $\tau = (\sqrt{5}+1)/2$. This irrational also gives the length scale for the transformations called inflations (deflations) that leave the tiling invariant, in which the
basic units of the tiling are redefined so as to give a Penrose
tiling on a larger (smaller) scale. These and many other fascinating properties of the Penrose tiling have been extensively studied in the literature ~\cite{levine}.   
This type of ordered structure can lead to complex physics, as shown by a large number of studies on electronic properties in this and 
other quasiperiodic models~\cite{sire,grimm}. Quasiperiodic quantum spin chains have also been the subject of many studies. The recent interest in quantum entanglement of spins has led for example, to investigation of one dimensional critical aperiodic systems \cite{igloi} showing that the entanglement entropy depends on the strength of the aperiodicity. Quantum effects are biggest in low dimensions and small spin value, while two is the smallest dimension for which T=0 order can occur. We therefore investigate the consequences of a quasiperiodic 
geometry in two dimensions for a Heisenberg $S=\frac{1}{2}$ antiferromagnet.


In an
antiferromagnet, quantum fluctuations around the Neel state lead to 
a reduction of the order parameter with respect to its classical
value, even at $T=0$. On bipartite Archimedean lattices, where all sites have the same value
of $z$, the staggered
magnetization is expected to increase with $z$, towards the classical value of $\frac{1}{2}$. This effect is easily explained within linear spin wave theory~\cite{ander}, and it is confirmed in a number of numerical calculations.
Thus for example, the order parameter on
the honeycomb lattice ($z=3$), $m_s \approx 0.235$~\cite{rieger}, is more strongly suppressed than on
the square lattice ($z=4$), where $m_s \approx  0.307$~\cite{sandvik}.

For inhomogeneous $ordered$ structures with more
than one value of $z$, it was recently argued that, contrarily to naive belief based on the preceding remarks, quantum fluctuations in the
ground state are typically $greater$ on sites with greater $z$~\cite{jag1}. 
Compared to the previous structures studied, the Penrose tiling is the most 
complex, with more local environments and more complex transformation rules than the quasiperiodic octagonal tiling. 
The ground state of the former has significantly stronger variations of the local order parameters as compared to the latter. The results show a strong decrease of onsite magnetization with $z$ for small $z$, followed by an upturn for larger $z$ -- a behavior we will explain by generalizing an argument presented in Ref.~\onlinecite{jag1}.

The ground state of the Penrose antiferromagnet can be described in terms of the local staggered magnetizations. We calculate these by two different methods: linearized spin wave (LSW) theory and quantum Monte Carlo (QMC). Although the real space distribution of the local staggered 
magnetization thus found is complex, a compact visualization of it is possible in ``perpendicular space", as will be explained
below. 

The model we consider is the nearest neighbor Heisenberg antiferromagnet
\begin{eqnarray}
\mathcal{H} = \sum_{\langle i,j\rangle} J \vec{S}_i\cdot \vec{S}_j,
 \label{ham}
\end{eqnarray}
where the sum is taken over pairs of linked sites and all bonds $J>0$ are of the
same strength. The site index $i$ takes values 1 to $N$, for the
finite size systems considered. The first type of systems we
consider are periodic approximants called Taylor approximants --
after their use in the description of the Taylor phases
of intermetallic compounds in the Al-Pd-Mn system~\cite{taylor} --
which allow for using periodic boundary conditions. These
approximants can be constructed in such as way as to obtain
sublattices of equal size, and we have considered four such systems, with
$N =96, 246, 644$ and 1686 sites. These approximants have defects as
compared to the infinite perfect tiling, but the relative number of
defects becomes negligible as $N$ increases. We also considered
finite pieces of the perfect Penrose tiling and find that
spin magnetizations in the interior of the finite sample are close to
those obtained for the Taylor approximants, showing their relative insensitivity to boundary conditions.

The model of Eq.~(\ref{ham}) is unfrustrated, and the ground state of this bipartite system breaks
the $SU(2)$ symmetry of $H$, with the order parameter being
the staggered magnetization $M_s = \sum_i \epsilon_i \langle
\vec{S}_i^z \rangle \equiv \sum_i m_{si}$, where $\epsilon_i =\pm 1$
depending on whether $i$ lies in sublattice A or B and
 $m_{si} = \vert \langle S^z_i \rangle \vert $ are the local
order parameters.

Within the quantum Monte Carlo (QMC) simulations, we obtain 
$m^2_{si}=\frac{3}{N}\sum_{j=1}^{N} \epsilon_i \epsilon_j \langle S^z_i
S^z_j \rangle$ from the spin-spin correlation
functions~\cite{wess1}. The QMC simulations were performed using the
stochastic series expansion method~\cite{sandvik} for the Taylor
approximants at temperatures chosen low enough to obtain ground
state properties of these finite systems~\cite{wess1}.

To obtain the spin wave Hamiltonian, one uses the Holstein-Primakoff boson
representation of $S^z$ on each sublattice in terms of the deviation
from the classical values of $\pm S$ , $ S_i^z = S - a_i^\dag a_i$ and $
S_j^z = -S + b_j^\dag b_j $, respectively~\cite{manou}. The $a_i$, $b_j$ $(i,j=1,...,N/2)$ and their adjoints, obey
appropriate bosonic commutation relations and correspond to the
sites of the A and B sublattices respectively. The spin raising and lowering operators
on the two sublattices are $S_i^+  = \sqrt{2S}\left(1 -
\frac{n_i}{2S}\right)^{\frac{1}{2}} a_i$ and $ S_j^+ =
\sqrt{2S}b_j^\dag \left(1 - \frac{n_j}{2S}\right)^{\frac{1}{2}}$, respectively.
After expanding to order $1/S$, the (LSW) Hamiltonian can be diagonalized
by a generalized Bogoliubov transformation~\cite{white}.
The ground state energy and $m_{si}$ can then be calculated from the  transformation matrix
(c.f. e.g. Ref.~\onlinecite{wess2}).
The LSW result for the ground state energy, extrapolated to the thermodynamic limit is $E_0/N=-0.643J$, and compares well to 
the QMC result, $E_0/N=-0.6529(1)J$.


\begin{figure}[t]
\begin{center}
\includegraphics[scale=0.7]{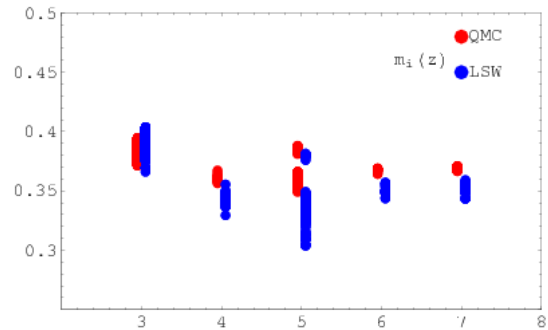}
\vspace{.2cm} \caption{(Color online) Local staggered magnetization plotted vs. coordination number $z$ as obtained by QMC (red) and by LSW theory (blue).} \label{qmc-sw.fig}
\end{center}
\end{figure}

Fig.~\ref{qmc-sw.fig} shows the values of $m_{si}$ plotted against
coordination number $z$ for the largest  approximant ($N=1686$) for both the LSW and QMC data. In comparison with the other known quasiperiodic structure, the octagonal tiling (see \cite{wess2}), the variations of the local order parameters are larger, making it possible to identify some of the trends more clearly. 
The values initially decrease with $z$,  but then
tend back upwards. There appears thus to be a minimum in
$m_s(z)$ at  $z=5$, the median $z$ value in this tiling
(Nb. on the infinite tiling as well as the approximants, the mean
value of $z$ is exactly 4). The average value of the magnetizations is also higher on the Penrose tiling, compared to the octagonal tiling, showing a suppression of quantum fluctuations due to greater structural complexity. 

Another noteworthy feature is the wide spread in the values for $z=5$.
This is related to the complex structural properties of the lattice, as there are three sets
of sites with $z=5$. The first set, which occurs  most
frequently, does not possess local five-fold symmetry and corresponds to the intermediate 
range of values of $m_{si}$. The two other sets of sites have a
five-fold symmetry and are at the centers of football-shaped
clusters (F) or star-shaped clusters (S). F sites correspond to the lowest $m_{si}$ values 
while the highest $m_{si}$ values are  obtained at the S sites. 

\begin{figure}[t] 
\begin{center}
\includegraphics[scale=0.45]{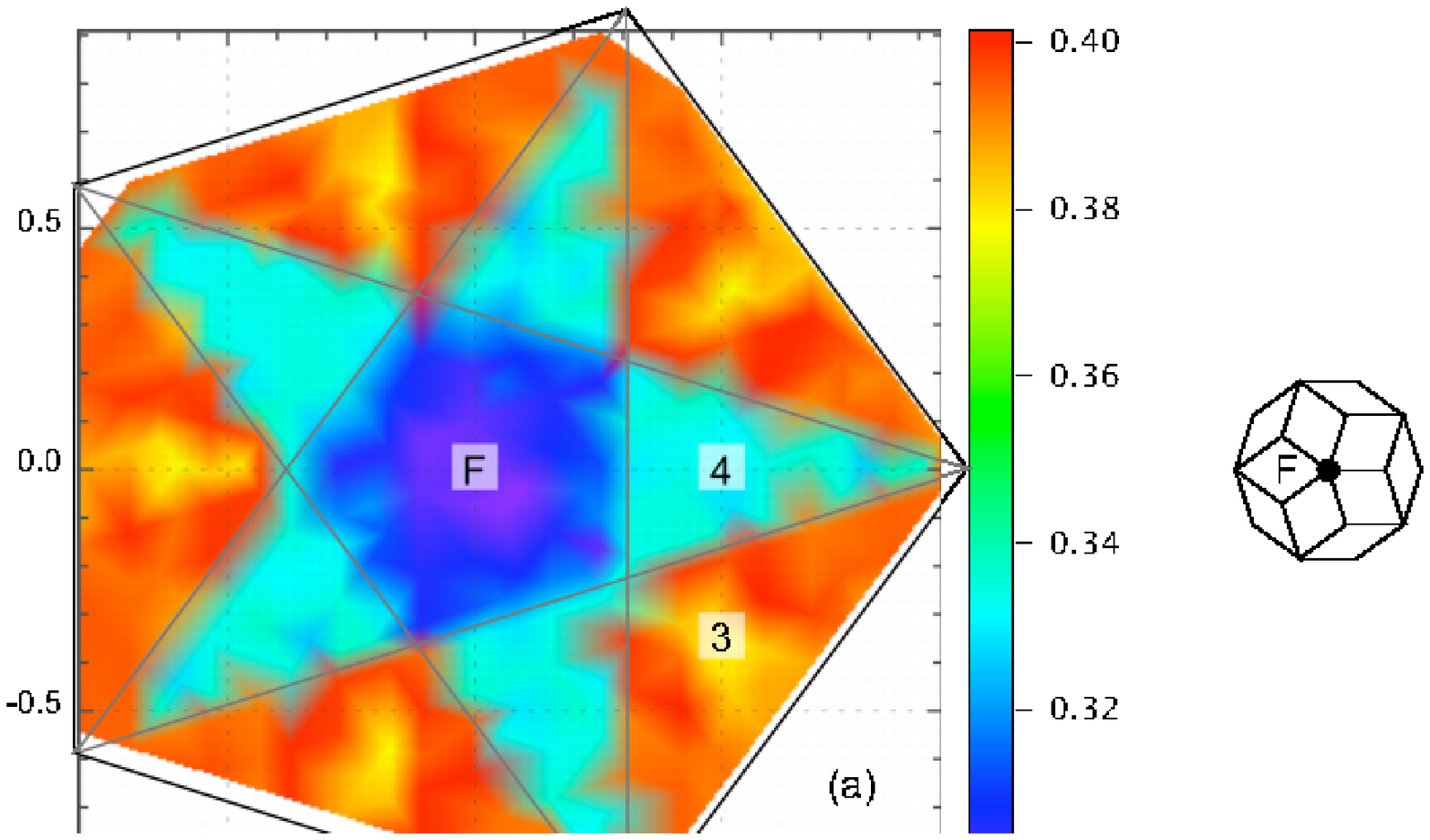}
\includegraphics[scale=0.45]{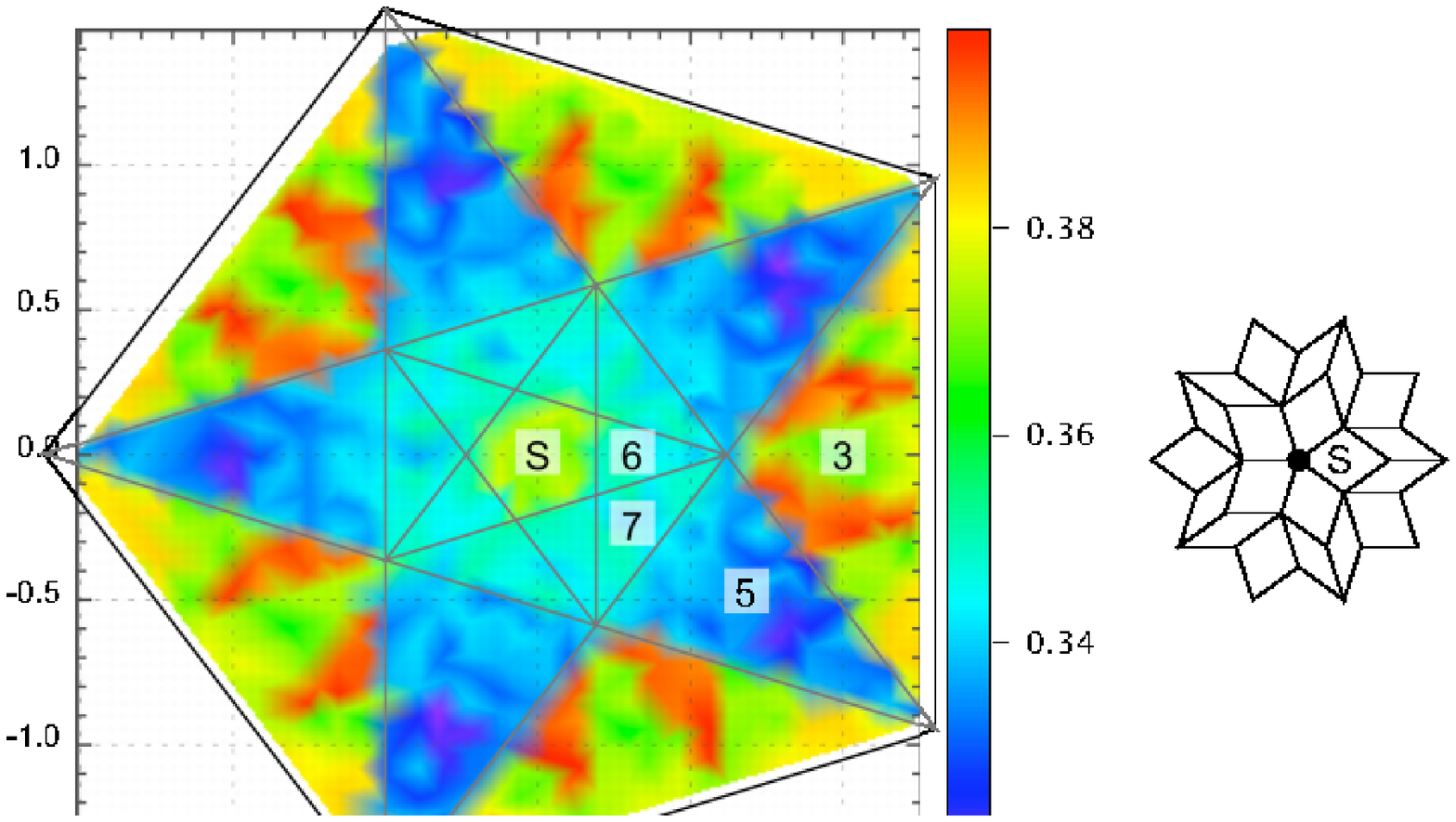}
\vspace{.2cm} \caption{(Color online) Two out of the four perpendicular space projected domains of the Penrose tiling,
with a color coding of the sites according to the value of the local staggered magnetization determined by linear spin wave theory.} \label{perpfig.fig}
\end{center}
\end{figure}

This local hierarchy in the magnetic structure on the Penrose tiling becomes evident in the
 ``perpendicular space" structural representation~\cite{levine}. The vertices of the Penrose
tiling can, in effect, be considered as the projection of vertices of a five dimensional cubic lattice onto the x-y (``physical") plane.
If those vertices are instead projected onto the
three remaining dimensions or ``perpendicular" space, one obtains dense packings of points lying on
four distinct pentagon-shaped plane regions. In this perpendicular space projection, sites having the same environment
map into the same subdomain of the selection windows (applied to a crystalline structure, the same operation
would lead to as many points as there are distinct environments, of which there are a finite number, contrarily to the quasicrystal).
The different domains are labeled in Fig.~\ref{perpfig.fig} by the
value of $z$ associated with each domain. In addition, the domains corresponding
to the sets of F and S sits are shown, along with
their appearance in real space.

Using a color map to represent the
local
 order parameters strengths, we obtain compact representations of the ground state 
 as in Fig.~\ref{perpfig.fig}, which thus shows the LSW magnetizations of
sites corresponding to two of the perpendicular space planes (the two others being
 identical upto rotations).
The points in the central star-shaped region of
Fig.~\ref{perpfig.fig}a correspond to the F sites, and have the
smallest staggered magnetizations. In Fig.~\ref{perpfig.fig}b the
central pentagon corresponding to the S sites, which have the highest staggered
magnetizations at $z=5$.

\begin{figure}[t]
\begin{center}
\includegraphics[scale=0.4]{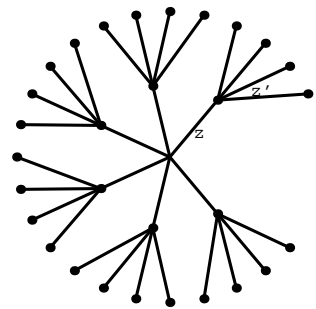}
\vspace{.2cm} \caption{A two-level Heisenberg star showing the central
spin, its $z$ nearest neighbors and $zz'$ next-nearest neighbors. In the example shown, $z=6$ and $z'=4$. }
\label{hstar2.fig}
\end{center}
\end{figure}

A simple model for the local staggered magnetization
considers a Heisenberg star cluster consisting of a central
spin coupled to $z$ neighboring spins \cite{jag1}. One considers
the external spins to be embedded in an infinite medium, so that
there is a finite net staggered magnetization. Carrying out the standard expansion
in boson operators, one then finds that the
onsite staggered magnetization of the central spin is lower than that
of the outer spins. This model, which takes
into account only the nearest neighbors is
inadequate to describe the non-monotonic dependence of magnetizations observed. 
We consider therefore a generalization to a two-level Heisenberg
star in order to investigate the effects of next-nearest neighbors
on the center spin magnetization. The cluster we consider is shown in Fig.~\ref{hstar2.fig}, where the
central site has $z$ nearest neighbors  and $zz'$
next-nearest neighbors. All the couplings (represented by the links in the figure) are taken equal, with $J>0$.

 The Hamiltonian of this cluster of
$1+z(1+z')$ spins can be diagonalized in linear spin wave theory,
with the following result for the central spin's staggered magnetization:
\bea m_s(z,z') = \frac{1}{2} -\frac{ zf_1^2(z,z')}{ f_2^2(z,z')- zf_1^2(z,z') - 4z'}, \label{hstarmag} \eea
where $f_{1(2)}= -z' \pm (2-z+\sqrt{4-4z +(z+z')^2})$. This yields a staggered magnetization that approaches the classical limit of 0.5 in the limit of large $z$ and/or $z'$. In addition, 
for fixed $z$ this function
$m_s(z,z')$ has a minimum for a value of $z'$ between $z-1$ and $z$. In other
words, the quantum fluctuations on the central site are largest when this site
and its neighbors have similar coordinations.

\begin{figure}[t]
\begin{center}
\includegraphics[scale=0.75]{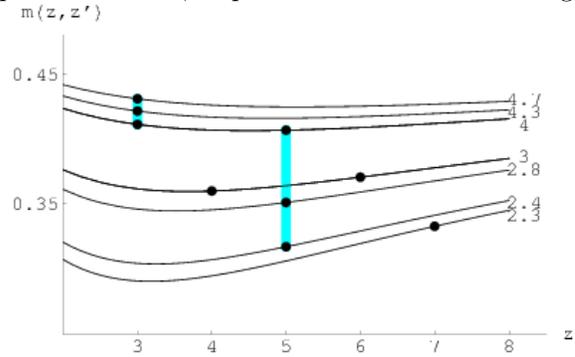}
\vspace{.2cm} \caption{(Color online) Staggered magnetization as predicted by Eq.~(\ref{hstarmag}) as a function of $z$ for different $z'$ values. The points indicate the value of z' computed (see text) for sites of the Penrose tiling.}\label{predict.fig}
\end{center}
\end{figure}

Turning now to the Penrose
tiling, effective values of $z'$ can be assigned for each site from counting
the number of its next-nearest neighbors. One finds that
sites of small $z$ have higher values of $z'$ (next nearest
neighbor number), with the opposite being true for sites of high
$z$. This means that the density of sites, in other words, does not have large local fluctuations on the Penrose tiling.
A single effective $z'$ is found for all the sites except for the values $z=3$ and $z=5$. For the $z=3$ sites,
we find $z' = 4,4.3$ and $4.7$, where the non-integral values result
from the fact that the clusters on the tiling do not have the regular tree
structure of the model shown in Fig. \ref{hstar2.fig}. This leads to a spread in the 
values of the local staggered magnetizations.
The generic $z=5$ sites correspond to $z'= 2.8$, while F and S sites have $z'=2.4$ and $4$, respectively. The resulting values for the $m_s(z,z')$ obtained
using Eq.~(\ref{hstarmag}) along with the values of $z$ and $z'$ for
each class of site are shown in
Fig.~\ref{predict.fig}.

The predictions of the simple
analytical model, which is based upon the number of nearest and next-nearest
neighbors only, agree qualitatively quite well with the
numerical results shown in Fig.~\ref{qmc-sw.fig} for most $z$. 
The complete description must of course include longer ranged structural differences, seen clearly in Figs.~\ref{perpfig.fig}: the domains of sites of a given coordination number are not
colored uniformly but are instead further separated into subdomains. 
The hierarchical invariance of the original structure, which has not been exploited in
these calculations (as was done in Ref. \onlinecite{jag2} using a renormalization group approach for the octagonal tiling) 
is expected to
lead to self-similarities in the order parameter distribution function. This analysis, which requires considering much bigger sample sizes, is left
for further investigations.

In conclusion, we have considered quantum fluctuations in the Penrose tiling, a two dimensional structure that has perfect long range structural order but with an infinite number of spin environments. The overall value of the staggered magnetization is higher than on the octagonal tiling, which is in turn higher than on the square lattice. This indicates a progressive suppression of quantum fluctuations in going from the periodic, to the simple quasiperiodic, and finally the more complex quasiperiodic structure. The geometry of the Penrose tiling leads to an antiferromagnetic ground state with extremely large variations
of the local staggered magnetization compared to other systems studied recently in this context.  The heirarchical symmetry present in the ground state is best seen in perpendicular space projections such as the ones shown in this paper. Finally, to explain our results, we present a two-level Heisenberg star argument showing that quantum fluctuations tend to be maximized when the site coordination number and the next nearest neighbor coordination numbers are closely matched in value.

\acknowledgments We would like to thank Boris Vacossin and Roderich Moessner for useful discussions, and Fran\c{c}ois Delyon for help with the figures. We thank HLRS Stuttgart and NIC J\"ulich for allocation of computing time.


\begin{references}

\bibitem{penref} R. Penrose, The Math. Intelligencer {\bf 2}, 32 (1979).

\bibitem{levine} P. J. Steinhardt and S. Ostlund in {\it The Physics of Quasicrystals}, World Scientific, Singapore 1987.

\bibitem{sire} C. Sire in {\it Lectures on Quasicrystals} eds. F. Hippert and D. Gratias, Les Editions de Physique, Les Ulis 1994.

\bibitem{grimm} U. Grimm and M. Schreiber in {\it Quasicrystals - Structure and Physical Properties}, ed. H.-R. Trebin (Wiley-VCH, Weinheim, 2003).

\bibitem{igloi} Ferenc Igloi, Robert Juhasz and Zoltan Zimboras, arXiv:cond-mat/0701527

\bibitem{ander} P. W. Anderson, Phys. Rev. {\bf 86},694 (1952); R. Kubo, Phys. Rev. {\bf 87}, 568 (1952).

\bibitem{rieger} Rieger \textit{et al.}, J. Phys. (Cond.Mat) {\bf 1} 1855 (1989).

\bibitem{sandvik} A. W. Sandvik, Phys. Rev. B {\bf 59}, R14157 (1999).

\bibitem{jag1} A. Jagannathan, R. Moessner and S. Wessel, Phys. Rev. B {\bf 74} 184410 (2006).

\bibitem{wess1} S. Wessel, A. Jagannathan and S. Haas,  Phys. Rev. Lett. {\bf 90}, 177205 (2003).

\bibitem{taylor} M. Duneau and M. Audier in {\it Lectures on Quasicrystals, Winter school Aussois Jan. 1994},  F. Hippert and D. Gratias Ed., Les Editions de Physique (1994).

\bibitem{manou} E. Manousakis, Rev. Mod. Phy. {\bf 63},1 (1991).

\bibitem{white} R. M. White, M. Sparks and I. Ortenburger, Phys. Rev. {\bf 139}, A450 (1965).

\bibitem{wess2} S. Wessel and I. Milat, Phys. Rev. B {\bf 71}, 104427 (2005).

\bibitem{jag2} A. Jagannathan, Phys. Rev. Lett. {\bf 92} 047202 (2004).


\end{references}
\end{document}